\documentclass[12pt]{article}
\usepackage{epsfig,amsmath}
\usepackage{axodraw}
\newcommand{\address}[1]{\\{\normalsize #1}}
\newcommand{\br}{\begin{eqnarray}}
\newcommand{\er}{\end{eqnarray}}
\newcommand{\bea}{\begin{eqnarray}}
\newcommand{\eea}{\end{eqnarray}}
\newcommand{\flin}[2]{\ArrowLine(#1)(#2)}
\newcommand{\wlin}[3]{\Photon(#1)(#2){2}{#3}}

\newcommand{\glun}[3]{\Gluon(#1)(#2){2}{#3}}
\newcommand{\lin}[2]{\Line(#1)(#2)}
\newcommand{\sof}{\SetOffset}
\def\PACS#1{\par\noindent\strut\kern18pt{\small\rm PACS numbers: #1}\par}

\def    \lsim {\raisebox{-3pt}{$\>\stackrel{<}{\scriptstyle\sim}\>$}}

\def \mh {\ifmmode M_H \else $M_H$\fi}

\newcommand     \sss            {\scriptscriptstyle}
\def \lambdahhh {\ifmmode \lambda_{\sss HHH}\else $\lambda_{\sss HHH}$\fi}
\def \lambdahhhzero {\ifmmode \lambda_{\sss HHH}^{(0)}\else
  $\lambda_{\sss HHH}^{(0)}$\fi}
\def \lambdahhhh {\ifmmode \lambda_{\sss HHHH}\else $\lambda_{\sss HHHH}$\fi}
\def \lambdahhhhzero {\ifmmode \lambda_{\sss HHHH}^{(0)}\else 
 $\lambda_{\sss HHHH}^{(0)}$\fi}
\begin{document}
\title{
\vspace*{-35mm}
{\normalsize
\begin{flushright}
UG-FT-173/04\\ CAFPE-43/04\\
December 2004\\
\end{flushright}
\vspace*{5mm}}
Physics beyond the Standard Model and its Minimal Supersymmetric 
extension at large colliders
\thanks{Presented by F. del Aguila at the final meeting of the European Network
``Physics at Colliders'', Montpellier, France, September 26--27, 2004.}
}

\author{F. del Aguila and R. Pittau 
\thanks{On leave of absence 
from Dipartimento di Fisica Teorica, Torino and INFN Sezione 
di Torino, Italy.}
\address{Departamento de F{\'\i}sica Te\'orica y del Cosmos and} 
\address{Centro Andaluz de F{\'\i}sica de Part{\'\i}culas Elementales
(CAFPE)}
\address{Universidad de Granada, E-18071 Granada, Spain}
}
\maketitle
\begin{abstract}
New large colliders will probe scales up to few TeV, indicating 
the way Nature has chosen to extend the Standard Model. 
We review alternative scenarios to the traditional Minimal 
Supersymmetric Standard Model: the little Higgs model, 
split supersymmetry and extra dimensional models with 
low energy signals.
\end{abstract}

%

\section{Introduction}

It is believed that the Standard Model (SM) is an effective 
low energy theory which has to be completed at some scale 
to include gravity. What aspects of this extension will 
show up first at large colliders will have to wait for 
new experimental data. In order to establish a departure 
from the SM predictions it will be probably necessary a new large 
striking signal. However, we will have to start carefully 
looking at the processes where the SM is less precisely known, 
like at the top or at the Higgs, or where clean signals can be 
observed, like lepton pair or monojet production. 

The most popular low energy extension of the SM is the 
Minimal Supersymmetric Standard Model (MSSM). Its main 
virtues are to make less severe the hierarchy problem: 
how to keep the Higgs boson mass 16 orders of 
magnitude lighter than the Planck scale 
$\sim 2.4 \times 10^{18}\ GeV$,
and to unify the gauge coupling constants of the three 
SM interactions near this scale. Its phenomenology 
is discussed in other contributions to these proceedings 
(see \cite{SUSYR0,SUSYR1,LCPWG,SUSYR2,PTC} 
for some recent reviews where this network has contributed). 
However, 
there are also other SM extensions which ameliorate 
its fine-tuning problems, and whose signatures can be also observable 
at large hadron colliders, like the Tevatron and the LHC, or 
at a large (International) Linear lepton Collider (ILC).  
In the following we review some of the implications 
of recent alternatives which may manifest at the top 
(Section 2), at the Higgs (Section 3), in 
lepton pair production with a large invariant mass (Section 4), 
and/or as an excess of monojets (Section 5).  

Two generic examples of alternative scenarios are the 
Little Higgs Model \cite{AHCG,AHCKN,LHM} and large Extra 
Dimensions \cite{ED,RS} (for recent (phenomelogical) reviews 
see \cite{EDR} (\cite{EDPR})). 
The former explotes our knowledge of four-dimensional 
gauge theories. 
It improves the hierachy problem related with 
the stability of the Higgs mass enlarging the SM to 
include a larger global symmetry at a higher scale 
$\sim TeV$, 
with a locally gauged subgroup containing the SM 
gauge group.
Its breaking above the $TeV$ is communicated to the 
SM Higgs doublet, which is a pseudo-Goldstone boson, 
through radiative corrections, and then, 
it is delayed by factors of $4\pi$ times coupling 
constants. 
The extra fields of its littlest version 
\cite{AHCKN} are a charged $\frac{2}{3}$ vector-like 
quark $T$, a complex scalar triplet $\phi$, and four 
new gauge bosons $A_H, W^{\pm}_H, Z_H$; all of them 
with masses near the $TeV$ scale, and eventually observable. 

On the other hand, the possibility of having observable 
large Extra spatial Dimensions raised in \cite{ED,RS} allows to 
reformulate the hierarchy problem. In this scenario 
there is only one fundamental scale of the order of 
the $TeV$, 
with some models being even Higgsless 
\cite{Higgsless0}.
The SM fields must mainly live in our 
four-dimensional brane, but can propagate, as does 
gravity, in the Extra Dimensions depending on the 
model assumptions. In this case new physics manifests 
as towers of Kaluza-Klein (KK) states, which, if light 
enough, would be observable at large colliders. 
Thus, we may want to know how to search for them.  
In practice, one can consider only the first 
massive states as we do below: new gauge bosons with the 
SM couplings but with heavier masses, 
new vector-like fermions with the SM quantum numbers 
and/or extra scalar replicas.   
  
This new framework also offers a new playground for 
dealing with other SM puzzles, as for instance, 
the necessity of accommodating the very different 
fermion masses observed in Nature. The paradigmatic
example is the very small neutrino masses. 
It was pointed out in \cite{EDN} that if 
Right-Handed (RH) neutrinos living in the 
bulk exist, their wave-function values on the brane 
where the standard Left-Handed (LH) neutrinos stay 
can be effectively so small as to explain the tiny 
Dirac neutrino masses observed; 
although to write definite realistic   
models, other mechanisms may be also required. The neutrino 
mass generation (see \cite{AF} for a review, and 
\cite{EDR} for the Extra Dimensional view), 
as the measured neutrino masses $m_\nu$, is 
not directly distinguishable at large colliders, because 
any related effect is suppressed by very small ratios 
$\sim \frac{m_\nu}{\sqrt s}$, where $\sqrt s$ is 
the collider center of mass energy. 
New neutrino effects at colliders must be associated 
to leptons with masses 
$\sim \sqrt s$ and relatively large mixings with 
the SM fermions. These new leptons (neutrinos), 
which may be also light KK modes, are discussed 
in other contribution to these proceedings 
\cite{M} (see also \cite{IT,BDNS,AAMM} and references 
there in).

Here we will be concerned with the (top) quark sector. 
The observed hierarchy between the light quarks and 
the top can be related to their localisation in 
the Extra Dimensions \cite{AHS}. 
Indeed, simple models with 
one extra dimension 
and quarks propagating in the bulk can reproduce the 
observed pattern of quark masses and mixings \cite{AS02}, 
and predict new vector-like quarks below the $TeV$ scale. 
Hence, a precise measurement of the top coupling $V_{tb}$ 
and the search of new heavy quarks at large colliders 
will constrain these models as well. 

Both SM alternatives,
the Little Higgs Model and the large Extra Dimensions, 
are more a class of models than 
definite complete theories. Their predictions are mainly 
estimates which may depend on unexpected new effects. 
This is the case in models with Extra Dimensions, 
for instance, of the so-called Brane Kinetic Terms (BKT). 
These are corrections to the kinetic terms of the bulk 
fields which are localised at the branes. 
They were first discussed in the context of 
gravity \cite{DGP}, but they are generated for bulk fields 
already at one loop in compactification with defects, 
as for instance in orbifolds \cite{GGH}. 
They are phenomenologically relevant \cite{CMS,CTW,APVSpBKT}, 
and deserve further theoretical study \cite{APVS}
(see \cite{APVS03} for a more complete set of references).
A review \cite{PV} can be found in these proceedings.  

Finally, as a last recent example of 
alternative to the MSSM it is worth to mention 
Split Supersymmetry 
\cite{AHDss,GRss}.
In this proposal squarks and sleptons are very heavy 
and can not even mediate the dominant decays of 
gauginos and higgsinos, which remain near the 
electroweak scale. The SM superpartners are thus 
split.
The plethora of new sparticles expected to be produced 
at forthcoming experiments in the MSSM gets reduced to 
some new (Majorana) fermions appearing, 
as we shall comment in Section 5, mainly as missing 
energy, plus a jet, for instance, in the gluino case.
A general discussion of the main phenomenological 
differences with traditional low-scale supersymmetry (SUSY), 
with emphasis in the cosmological aspects, 
is given in \cite{AHDGR}.

\section{Signals at the top}

Large colliders will be top factories. In particular the LHC 
with a cross section of $860\ pb$ will produce 
tens of millions of top pairs
\cite{TopB} (see also \cite{TopG,TopAS} in these proceedings),
and a large number of pairs of any new heavy quark 
which may exist with a mass up to several $TeV$.  
Single top production will be down by more than an order of 
magnitude, but it will give the best measurement of the 
top coupling to the $W$, $V_{tb}$, 
with an expected precision of $5 \%$ \cite{SSW}. The top 
coupling to the $Z$, $X_{tt}$, will be measured with a 
better precision at ILC, $2 \%$ \cite{LCPWG}. 
These two measurements could give evidence of new physics 
if different from their SM values 
$V^{\rm SM}_{tb} = 0.999$ and $X^{\rm SM}_{tt} = 1$, respectively. 
The most probable explanation of such a large deviation 
$V_{tb} < 0.96$ from the SM predictions would be the top 
mixing with other vector-like quark of charge 
$\frac{2}{3}$, $T$ \cite{AASM}. 
As emphasized in the Introduction, 
alternative SM extensions like the Little Higgs Model 
or Extra Dimensions predict such new vector-like quarks 
\cite{AHCKN,AS02}. 

\begin{figure}[]
\begin{center}
\epsfig{file=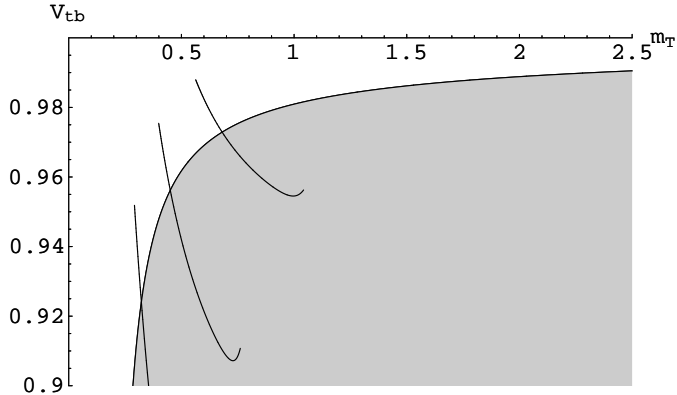,width=9cm} 
\caption{Excluded region (shadowed) in the $m_T - V_{tb}$ plane 
defined by requiring the T parameter to deviate by more than 
$3 \sigma$ from its experimental value. Lines show the $m_T - V_{tb}$ 
values for models with an extra dimension parameterising the 
orbifold $\frac{S^1}{Z_2}$ and radius $R = \frac{1}{M_c}$ with 
$M_c = 0.5,\ 0.7,\ 1\ TeV$, from left to right. The points along 
the lines correspond to growing BKT coefficients, from 0 to $20 R$ 
from right to left.} 
\label{Fig1}
\end{center}
\end{figure}

In the simplest case of only one effective light $T$ 
quark the $t - T$ quark system is parameterised by 
the two quark masses $m_t < m_T$ and its mixing angle 
\begin{equation}
V_{Tb} = \sqrt {1-V_{tb}^2} \ .
\end{equation}
($V^{\rm SM}_{tb}$ can be taken equal to 1). 
The corresponding mass matrix reads
\begin{equation}
\left( 
\begin{matrix} m_t & V_{Tb}m_T \cr
                0 & m_T 
\end{matrix}
\right)\ ,
\end{equation}
where the RH quarks (columns) have been rotated and the 
field phases chosen conveniently, and 
$V_{Tb}  \sim {\cal O} (0.1) \ll 1$ is 
a small expansion parameter. 
The physics of this SM addition has been often discussed in 
the literature \cite{VLTquark}. We follow the analysis in \cite{AS}.
The main constraint from precise electroweak data results from the 
{\it oblique} parameter \cite{LEPEWWG}
\begin{equation}
{\rm T} = 0.12 \pm 0.10\ .
\label{Texp}
\end{equation}
This latest value differs significatively from the value 
quoted in \cite{PDB} but it has been obtained fixing 
the parameter ${\rm U} = 0$, 
what seems more adequate for this type of SM extensions.
{\rm T} is a function of the $T$ mass and of its couplings to 
the $W$ and to the $Z$ squared,
$|V_{Tb}|^2$ and 
\begin{equation}
|X_{Tt}|^2 = |V_{Tb}V_{tb}|^2 = |V_{Tb}|^2 (1-|V_{Tb}|^2)\ ,
\end{equation}
respectively. Its approximate expression can be written
($m_t^2 \ll m_T^2$)
\begin{equation}
T = \frac{N_c}{16 \pi \sin ^2 \theta_W \cos ^2 \theta_W}
|V_{Tb}|^2 [-18.4+7.8\log \frac{m_T^2}{M_Z^2}+O(\frac{M_Z^2}{m_T^2})]\ ,
\label{aT}
\end{equation}
where $N_c = 3$ is the number of colours, and 
$\sin ^2\theta _W = 0.231$ and $m_t$ taken equal to $176\ GeV$ are the 
{$\overline {\rm MS}$}
values of the electroweak mixing angle and the top mass at the 
$M_Z = 91.2\ GeV$ scale, respectively.
In Fig. \ref{Fig1} 
%
%
we plot the experimentally excluded region (shadowed) 
of the $m_T - V_{tb}$ plane \cite{APVSpBKT}. 
The limiting curve defines the 3 standard deviation bound 
${\rm T} \le 0.29$, 
and only masses and mixings on the top-left part are allowed. 

In the Little Higgs Model \cite{LHM}
\begin{equation}
V_{Tb} = \frac{\lambda _1}{\lambda _2} \frac{m_t}{m_T}\;\;\;\;  {\rm and}
\;\;\;\;  V_{tb} = 1 - \frac{1}{2}
\frac{\lambda _1^2}{\lambda _2^2} \frac{m_t^2}{m_T^2}\ ,
\label{LHMVTb}
\end{equation}
with $\lambda _{1,2}$ two couplings of ${\cal O} (1)$ satisfying 
\begin{equation}
\frac{1}{\lambda _1^2} + \frac{1}{\lambda _2^2} 
\simeq 
\frac{v^2}{m_t^2}\ ,
\end{equation}
where $v = 246\ GeV$ is the Higgs doublet vacuum expectation value, 
and $m_T$ 
naturally ranging from 1 up to 5 $TeV$, but otherwise arbitrary.
(In more restrictive models of this class the non-observation 
of an excess of lepton pairs at Tevatron banishes $m_T$ 
to higher values \cite{LHMR}.)
Then Eq. (\ref{aT}) and the 3 $\sigma$ limit on T above further 
constrain 
\begin{equation}
\frac{\lambda_1}{\lambda _2} 
\le 
\frac{m_T (TeV)}{\sqrt {0.681 + 0.560 \log m_T (TeV)}}
\ ,
\label{LHMl}
\end{equation}
for large $m_T$.
Hence, already for $m_T = 1\ TeV$, $V_{tb} > 0.96$, and no 
distinction between the Little Higgs Model and the SM can 
be made with the foreseen precision for the $V_{tb}$ 
measurement. 
On the other hand, models with Extra Dimensions and 
multilocalisation can accommodate smaller $m_T$ and $V_{tb}$ 
values \cite{AS02}. In these models they fix 
two free parameters: the five-dimensional top mass and 
the position of the intermediate brane. 
The allowed $m_T - V_{tb}$ region can be also reached 
in the simplest models without multilocalisation 
if relatively large BKT are included \cite{APVSpBKT}. 
In Fig. \ref{Fig1} we draw  
lines of constant compactification scale $M_c$ and growing 
BKT from right to left. A model with excluded 
values in the shadowed region can 
predict allowed $m_T$ masses and $V_{tb}$ couplings 
increasing (moving from right to left along the lines) 
the BKT coefficient.  
In summary, a new vector-like $T$ quark with a large 
mixing with the top could be the first signal of 
Extra Dimensions.

\begin{figure}[t]
\begin{center}
\unitlength=1cm
\begin{picture}(10,7)
\put(-10.5,-29){\includegraphics[width=30cm,angle=0]{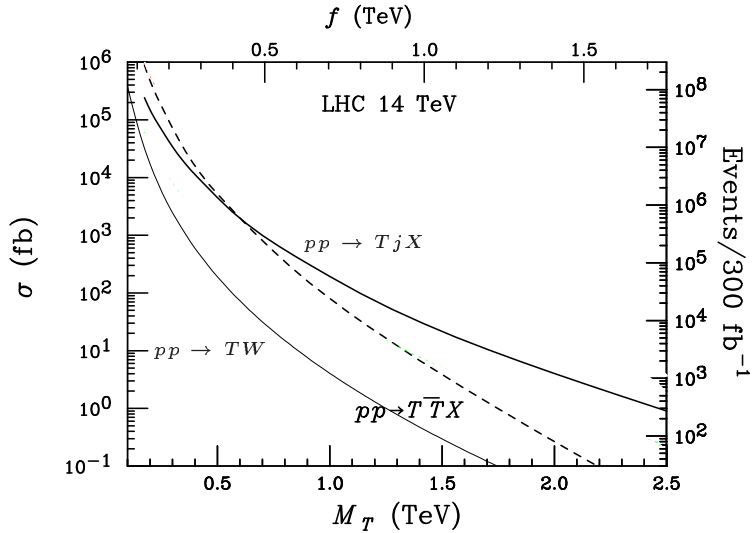}}
\end{picture}
\end{center}
\caption{
Cross sections for different $T$ production processes at LHC:
$pp\rightarrow T \bar T X$ (dashed), $pp\rightarrow T j X$ 
(thick solid) and $pp\rightarrow T W X$ (thin solid).}
\label{Fig2}
\end{figure}

\begin{figure}[]
\begin{center}
\begin{picture}(200,160)
\SetWidth{1}
\Text(43,62)[b]{(b)}
\Text(153,62)[b]{(c)}
\SetScale{0.62}
\sof(-10,20)
\lin{40,0}{87.5,0} \flin{40,50}{87.5,50}
\wlin{87.5,0}{87.5,50}{6}
\flin{87.5,50}{135,50} \lin{87.5,0}{135,0}
\Text(22.94,31)[br]{\tiny $b$}
\Text(21,-2.7)[lt]{{\tiny (-)}}
\Text(22.94,-9)[lt]{\tiny $q$}
\Text(58.59,17)[l]{\tiny $W$}
\Text(85.56,31)[lb]{\tiny $T$}
\Text(85.56, -2.8)[lt]{{\tiny (-)}}
\Text(87.5,-7)[lt]{{\tiny ${{q}^{~\prime}}$}}
\sof(100,20)
\flin{40,0}{65,25} \flin{65,25}{40,50}
\wlin{65,25}{110,25}{6}
\flin{110,25}{135,50} \flin{135,0}{110,25}
\Text(22.94,31)[br]{{\tiny ${\bar q}^\prime$}}
\Text(22.94,0)[tr]{{\tiny $q$}}
\Text(54.25,18.6)[b]{{\tiny $W$}}
\Text(85.56,31)[lb]{{\tiny $T$}}
\Text(85.56,0)[lt]{{\tiny $\bar b$}}
\sof(-80,100)
\glun{40,0}{87.5,0}{7} \glun{87.5,50}{40,50}{7}
\flin{87.5,0}{87.5,50}
\flin{87.5,50}{135,50} \flin{135,0}{87.5,0}
\Text(22.94,31)[br]{\tiny $g$}
\Text(22.94,0)[tr]{\tiny $g$}
\Text(58.59,15.5)[l]{\tiny $T$}
\Text(85.56,31)[lb]{\tiny $T$}
\Text(85.56,0)[lt]{\tiny $\bar{T}$}
\sof(0,100)
\glun{40,50}{87.5,0}{9}
\glun{40,0}{87.5,50}{9}
\flin{87.5,0}{87.5,50}
\flin{87.5,50}{135,50} \flin{135,0}{87.5,0}
\Text(22.94,31)[br]{\tiny $g$}
\Text(22.94,0)[tr]{\tiny $g$}
\Text(58.59,15.5)[l]{\tiny $T$}
\Text(85.56,31)[lb]{\tiny $T$}
\Text(85.56,0)[lt]{\tiny $\bar T$}
\sof(85,100)
\Text(11,42)[b]{(a)}
\glun{40,0}{65,25}{5} \glun{65,25}{40,50}{5}
\glun{65,25}{110,25}{7}
\flin{110,25}{135,50} \flin{135,0}{110,25}
\Text(22.94,31)[br]{\tiny $g$}
\Text(22.94,0)[tr]{\tiny $g$}
\Text(54.25,19)[b]{\tiny $g$}
\Text(85.56,31)[lb]{\tiny $T$}
\Text(85.56,0)[lt]{\tiny $\bar T$}
\sof(170,100)
\flin{40,0}{65,25} \flin{65,25}{40,50}
\glun{65,25}{110,25}{7}
\flin{110,25}{135,50} \flin{135,0}{110,25}
\Text(22.94,31)[br]{\tiny $\bar q$}
\Text(22.94,0)[tr]{\tiny $q$}
\Text(54.25,19)[b]{\tiny $g$}
\Text(85.56,31)[lb]{\tiny $T$}
\Text(85.56,0)[lt]{\tiny $\bar T$}
\end{picture}
\end{center}
\caption{
(a): Diagrams for ${T \bar T}$ production. (b) and (c): $t$- and
$s$- channel contributions, respectively, to $T j$ production.}
\label{Fig3}
\end{figure}
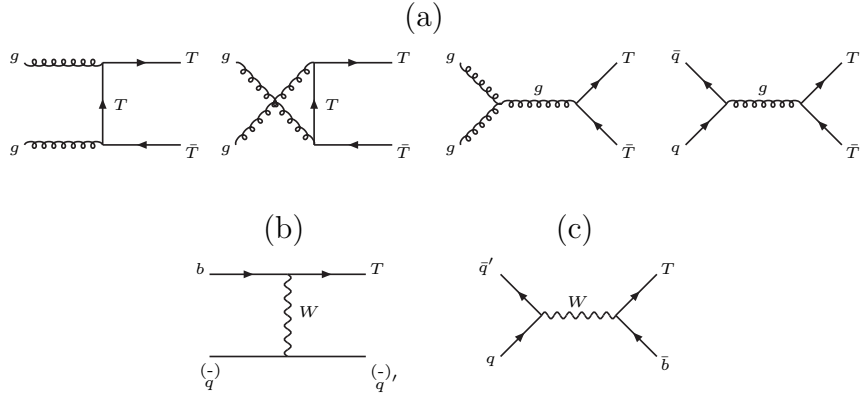

\begin{figure}[h]
\begin{center}
\begin{picture}(200,80)
\SetWidth{1}
\sof(-80,20)
\Text(135,42)[b]{(a)}
\Text(220,42)[b]{(b)}
\SetScale{0.62}
\sof(0,20)
\flin{40,0}{65,25} \glun{65,25}{40,50}{5}
\flin{65,25}{110,25}
\flin{110,25}{135,50} \wlin{110,25}{135,0}{5}
\Text(22.94,31)[br]{{\tiny $g$}}
\Text(22.94,0)[tr]{{\tiny $b$}}
\Text(54.25,18.6)[b]{{\tiny $b$ }}
\Text(85.56,31)[lb]{{\tiny $T$ }}
\Text(85.56,0)[lt]{{\tiny $W$ }}
\sof(85,20)
\flin{40,0}{87.5,0} \glun{87.5,50}{40,50}{7}
\flin{87.5,0}{87.5,50}
\flin{87.5,50}{135,50} \wlin{87.5,0}{135,0}{6}
\Text(22.94,31)[br]{{\tiny $g$}}
\Text(22.94,0)[tr]{{\tiny $b$}}
\Text(58.59,15.5)[l]{{\tiny $T$}}
\Text(85.56,31)[lb]{{\tiny $T$}}
\Text(85.56,0)[lt]{{\tiny $W$}}
\end{picture}
\end{center}
\caption{
$s$-channel (a) and $t$-channel (b)
$T W$ production diagrams.}
\label{Fig4}
\end{figure}
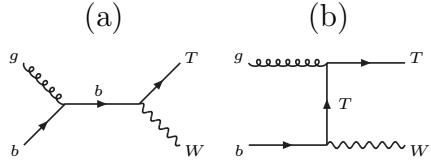

Even if the mixing of such a new quark is too small to be 
observed, it could still be produced at LHC if its 
mass is up to few $TeV$. In Fig. \ref{Fig2}
%
%
we plot the cross sections for different 
$T$ production processes at LHC \cite{LHM,PTC}. 
$T\bar T$ production is fixed by QCD and its cross section 
decreases faster with $m_T$ because the $t$-channel interchanges 
the $T$ quark and the $s$-channel is relatively suppressed at 
high energy (Fig. \ref{Fig3}(a)).  
%
%
$Tj$ production is model dependent. In Fig. \ref{Fig3}(b) we 
show the dominant process. The cross section in Fig. \ref{Fig2} 
corresponds to $V_{Tb} = \frac{m_t}{m_T}$. In the Little Higgs 
Model a ratio $\frac{\lambda _1}{\lambda _2} = \frac{1}{2}$ 
in Eq. (\ref{LHMVTb}) would stand for a factor 4 smaller cross 
section. Larger ratios are constrained by Eq. (\ref{LHMl}). 
The upper variable $f$ in this Figure is the parameter fixing 
the scale of new physics. The plotted values correspond to 
the lower limit $f \lsim \frac{v}{2m_t} m_T$. 
Models with Extra Dimensions can show a similar 
$V_{Tb}$ dependance \cite{AS02}. The $s$-channel (Fig. \ref{Fig3}(c)) 
is very much suppressed, because although it involves the same 
vertices as the $t$-channel, it has the high energy $s$-channel 
suppression relative to $t$-channel contribution. 
We also plot in Fig. \ref{Fig2} the $T W$ cross section. 
The corresponding diagrams are depicted in Fig. \ref{Fig4}. 
%
%
Although there is the enhancement of a strong coupling vertex and 
an initial gluon relative to the $T j$ contributions, the exchange 
of a $T$ quark in the $t$-channel diagram (Fig. \ref{Fig4}(b)) and 
the $s$-channel suppression at high energy (Fig. \ref{Fig4}(a)) reduce 
this cross section below the other processes.
However, to decide about the best signal one also must wonder 
about the background. In Table \ref{Table1} we give an estimate 
for the three cases.
\begin{table}[htb]
\begin{center}
\begin{tabular}{l|ccc}
$T$ production process & $pp\rightarrow T\bar T X$ & 
$pp\rightarrow T j X$ & $pp\rightarrow T W X$ \\
\hline
\hline
SM background 
 & $pp\rightarrow Z Z W W jj X$ & $pp\rightarrow Z W jj X$ & 
$pp\rightarrow Z W W j X$ \\
Estimate & $4\times 10^{-4}\ pb$ & $6\ pb$ & $8\times 10^{-2}\ pb$ \\
\end{tabular}
\end{center}
\caption{SM background estimates for different $T$ production processes 
evaluated with {\tt ALPGEN} \cite{alpgen}.}
\label{Table1}
\end{table}
Jets are required to have a transverse momentum $p_t^j > 40\ GeV$, 
a pseudorapidity $|\eta _j| < 2.5$ and a separation   
$\Delta r (jj) > 0.7$.
In all cases we assume that $T$ decays into $Z t$, what typically 
does $\frac{1}{4}$ of the time \cite{LHM,AAKV}, for it gives 
the cleanest signal. 
The largest $T$ cross section, $T j$ production, has also the largest 
background, but it can be reduced by a large factor, as also can the 
other backgrounds, requiring the $t \rightarrow W b$ and the 
$T \rightarrow Z t$ reconstruction (cutting in the corresponding 
invariant mass distributions). Cross sections must be multiplied 
in all cases by the corresponding branching ratios when looking at 
specific channels.

\section{A non-standard Higgs}

The SM Higgs physics as well as the SUSY and the
Two-Higgs Doublet Model (2HDM) \cite{2HDM} 
(see also \cite{BS}) Higgs sector are reviewed 
in other contributions to these proceedings \cite{MK,GD}.  
Here we concentrate on the types of models discussed 
in the Introduction.

The complete reconstruction of the Higgs potential necessarily 
requires the measurement of the Higgs self-couplings. These 
include a trilinear and a quartic interaction, parameterised 
by the coupling constants \lambdahhh\ and \lambdahhhh, which 
in the SM take the values
\begin{equation}
\label{lambdas}
\lambdahhhzero=-3\, \frac{M_H^2}{v} \; , \quad \lambdahhhhzero=-3 \,
\frac{M_H^2}{v^2}\ ,
\end{equation}
where $M_H$ is the Higgs mass.
A direct measurement of \lambdahhh\ can be obtained, both at LHC and ILC,
via the detection of Higgs boson pairs, while, due to the vanishingly
small 3-Higgs production cross section, there is no hope for
a measure of  $\lambdahhhh$.
Several models predict sizeable departures
of \lambdahhh\ from its SM value.
For example, a scan over the parameter space of the
2HDM
shows that values of
$r \equiv\lambdahhh/\lambdahhhzero$
such as $-15 < r < 15$ are quite possible \cite{higgsself}.

A second example is the Little Higgs Model \cite{LHM},  
with
\begin{eqnarray}
\label{eq:deltar}
1 \le r 
\simeq 1+
\delta_r\,~~~{\rm and}~~
\delta_r = 
\frac{\lambda ^2_{h\phi h}}
{\lambda_{\phi ^2} \lambdahhhzero} 
\ ,
\end{eqnarray}
where $\lambda _{h\phi h}$ and $\lambda _{\phi ^2}$ are 
coefficients of the enlarged Higgs potential after 
the spontaneous breaking of the assumed global symmetry. 
Besides the usual complex scalar doublet $h$, 
this Littlest Higgs Model potential contains an additional 
complex scalar triplet $\phi$. 
Then, the quartic Higgs doublet self-coupling $\lambda _{h^4}$, 
which also gives \lambdahhh\ , is not any more only fixed in terms
of the Higgs mass and $v$, but of this ratio of Higgs potential 
couplings function of the fundamental parameters of the model: 
gauge and Yukawa couplings, and loop coefficients.

\begin{figure}[!t]
\begin{center}
\epsfig{file=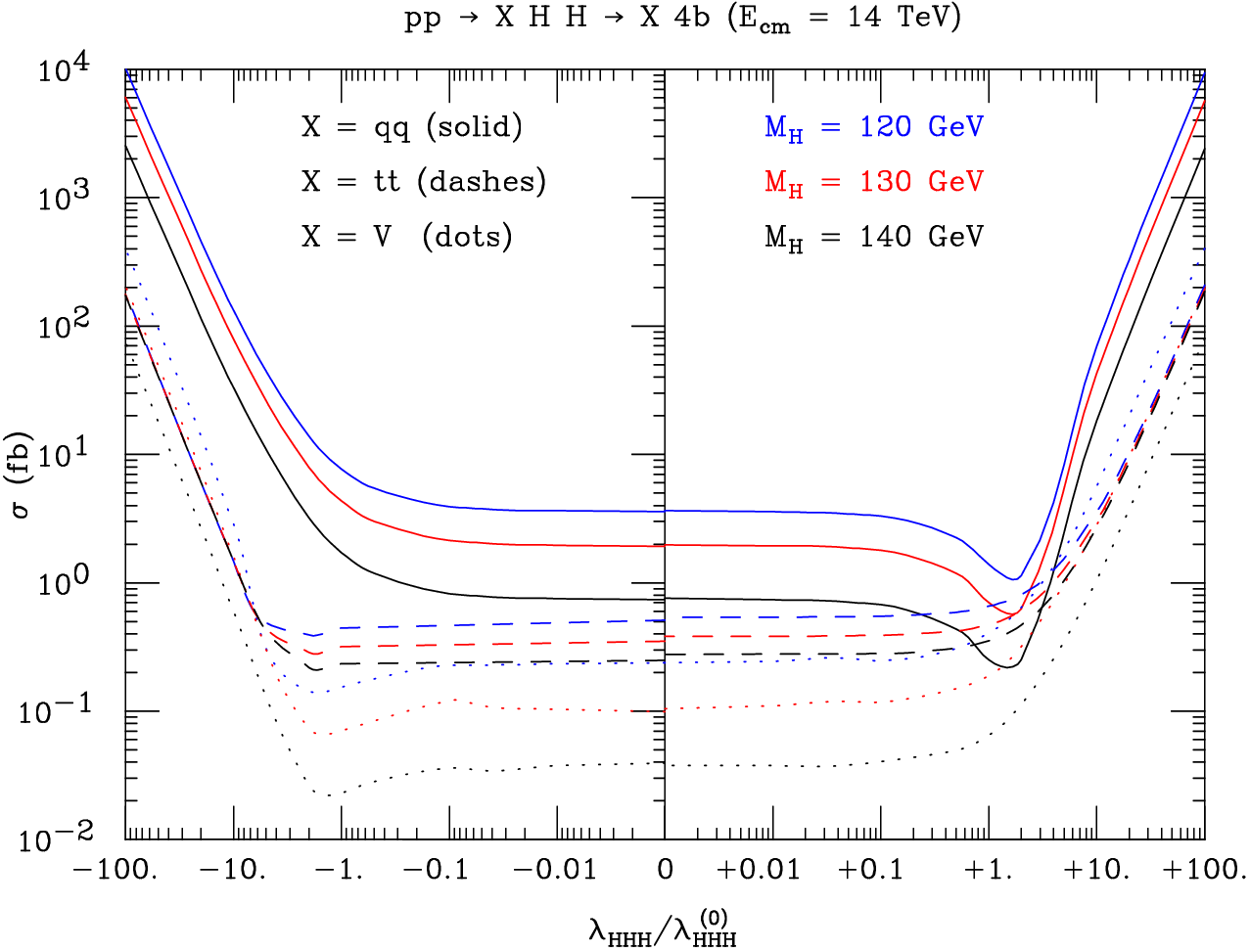,width=0.5\textwidth,angle=90}
\DashLine(-49,21.6)(-49,165.5){4}
\SetWidth{0.2}
\DashLine(-33.1,21.6)(-33.1,165.5){2}
\end{center}
\caption{\small Dependence of the cross sections for the
three processes
$q q^{(')}\to q q^{(')} {HH}$, $gg,q\bar{q}\to t\bar t HH$ and
$q\bar{q}^{(')} \to V {HH}$
on $r=\lambdahhh/\lambdahhhzero$,
in correspondence of three values of the Higgs mass.}
\label{fig:lambdas}
\end{figure}

In Fig. \ref{fig:lambdas}
%
%
we show the effect, at LHC,
of varying \lambdahhh\ independently
of any other SM parameter \cite{higgsself}.
Notice that the cross sections do not vanish at zero values of
\lambdahhh\ because of the presence of diagrams where the
two Higgses are radiated independently, with strength proportional to
the Yukawa or gauge couplings.
The two vertical lines correspond to the limits given in Eq.
(\ref{eq:deltar}) with $\delta_r= 5.1$.
The region at the right of the thin dashed line is accessible
for a $3\ \sigma$ discovery at LHC when $M_{H} \sim 120\ GeV$.
Such a discovery limit has been computed with
a detailed signal-to-background analysis performed with
{\tt ALPGEN} \cite{alpgen}.
This measurement with a large deviation from the SM prediction 
would further constrain this class of models.

Finally, the Higgs decay rates also show a dependance 
on the model, especially the induced decays: 
$H\rightarrow g g, \gamma \gamma$. 
This variation is expected to be relatively small 
in the Little Higgs Model, and difficult to observed at 
LHC or ILC \cite{LHMD}.

In models with Extra Dimensions the Higgs can mix with 
the new singlet scalars associated to the 
background metric. For instance, this mixing with the radion in the 
Randall-Sundrum model \cite{RS} also modifies the 
Higgs mass and decay rates. The universal coupling of the metric 
to the energy-momentum tensor gives no variation 
in the Higgs branching ratios into massive states
but modifies the total and partial decay rates, as 
well as the induced decay $H\rightarrow g g$ 
branching fraction \cite{radmix}. These modifications can 
be larger if the model has a non-minimal Higgs sector, 
as in the case of the 2HDM \cite{PTC}. 

\section{Large di-lepton signals}

Lepton pair production is expected to be the main 
production channel to discover and to study new gauge 
bosons at LHC and ILC (see \cite{A} for a review).
An excess of lepton pairs with a large invariant mass 
can be also the signature of a tower of KK gravitons 
\cite{EDPR}.
As a matter of fact, the large variety of models which  
predict new contributions to this final state 
will require a detailed study of this signal, and 
of other channels, to characterise its origin.
A resonant peak periodically repeated will point 
out to Extra Dimensions, but the mass of the first 
resonance will constrain the type of model. 
For instance, if relatively heavy, it will disfavour  
universal Extra Dimensions \cite{UED}.
On the other hand, its angular distribution will 
determine the spin assignment, which can be measured 
at the LHC, as well as at the ILC \cite{DHRed,OPP}.

In any case a detailed analysis of this final state 
together with other related channels, like 
$W$ pair production or even quark pair production, 
shall help to distinguish between models, for example,  
establishing its Higgsless character \cite{Higgsless1}.
(The simplest versions of Higgsless models 
\cite{Higgsless0} give 
a too large ${\rm S}$ {\it oblique} parameter 
\cite{Higgsless1,Higgsless2}, 
but this could be cured \cite{Higgsless3}.)

The Little Higgs Model also predicts new heavy gauge 
bosons with characteristic decay rates \cite{LHM}.
For a review of the phenomenology of SM gauge extensions, 
like for instance those based on $E_6$, 
see \cite{L} and references there in.

\section{Jets beyond the SM}

As in the case of di-leptons, missing energy is 
a signal of many SM extensions, supersymmetric or not. 
For example, any new neutral gauge boson coupling 
to quarks and neutrinos, as those predicted by the 
Little Higgs Model \cite{LHM}, would increase the 
SM background for mono-jets. 
In models with Extra Dimensions 
KK gravitons which escape detection can be also produced 
with a jet \cite{SUSYR1}. 

This signature is also characteristic of  
SUSY models. 
The most recent alternative to the MSSM, Split 
SUSY, can manifest as a mono-jet excess if the 
gluinos, which are assumed to be relatively light, 
hadronise little \cite{HLMR}. 
For a general discussion of the phenomenology of 
this new scenario see 
\cite{AHDGR}.

\section{Conclusions}

Alternatives to the SM beyond the MSSM like the Little Higgs 
Model, Models with Extra Dimensions (Higgsless or not) 
or Split SUSY offer a distintive solution to the hierarchy 
problem, and then new physics at the $TeV$ scale which can 
be eventually observed at large colliders. 
We have reviewed some of their implications for top and 
Higgs physics, and in di-lepton and monojet production. 
These two channels are the signature of many SM extensions 
and we have essentially referred to the literature.
In Table \ref{Table2} we gather the signal and collider 
where to test each class of models. 
\begin{table}[htb]
\begin{center}
\begin{tabular}{ccc}
\hline
\hline
Signal & Collider & New physics \\
\hline
\hline
Small $V_{tb}\ (X_{tt})$ & LHC (ILC) 
& Extra dimensions \\
\hline
{\begin{tabular}{c}
New $T$ quark \\ 
of charge $\frac{2}{3}$ \\
\end{tabular}} 
& LHC & 
{\begin{tabular}{c}
Extra dimensions \\ 
Little Higgs model \\
\end{tabular}}  \\
\hline
{\begin{tabular}{c}
Non-standard trilinear \\ 
Higgs coupling \\
\end{tabular}} 
& LHC & 
{\begin{tabular}{c}
Extra dimensions \\ 
Little Higgs model \\
\end{tabular}} \\
\hline
Higgs decay rates  
& LHC / ILC & 
{\begin{tabular}{c}
Extra dimensions \\ 
Little Higgs model \\
\end{tabular}} \\
\hline
{\begin{tabular}{c}
Large di-lepton \\ 
production \\
\end{tabular}} 
& LHC / ILC & 
{\begin{tabular}{c}
Extra dimensions \\ 
Little Higgs model \\
\end{tabular}} \\
\hline
{\begin{tabular}{c}
Non-standard jet \\ 
cross sections \\
\end{tabular}} 
& LHC & 
{\begin{tabular}{c}
Extra dimensions \\ 
Little Higgs model \\
Split SUSY
\end{tabular}} \\
\hline
\hline
\end{tabular}
\end{center}
\caption{Signals of alternative 
SM extensions at large colliders.}
\label{Table2}
\end{table}
See 
\cite{PTC}
for a recent review of these alternatives.

\vskip .5cm

We thank J.A. Aguilar-Saavedra, A. Mart{\'\i}nez de la Ossa, 
D. Meloni, M. P\'erez-Victoria and J. Santiago for assistance 
in preparing this review.
This work has been supported in part by MCYT under contract
FPA2003-09298-C02-01, by Junta de Andaluc\'{\i}a group FQM 101, by the
European Community's Human Potential Programme under contract
HPRN-CT-2000-00149 Physics at Colliders. R.Pittau thanks the financial 
support of MEC under contract SAB2002-0207.

\end{document}